\documentclass[12pt]{article}
\usepackage[english]{babel}
\usepackage[a4paper,top=2cm,bottom=2cm,left=3cm,right=3cm,marginparwidth=1.75cm]{geometry}
\usepackage[T1]{fontenc}
\usepackage{listings}

\usepackage{amsmath}
\usepackage{graphicx}
\usepackage[colorlinks=true, allcolors=blue]{hyperref}
\usepackage{authblk}
\usepackage{enumitem}
\usepackage[normalem]{ulem}
\usepackage{color, soul}
\usepackage{booktabs}

\newbox{\myorcidaffilbox}
\sbox{\myorcidaffilbox}{\large\includegraphics[height=1.7ex]{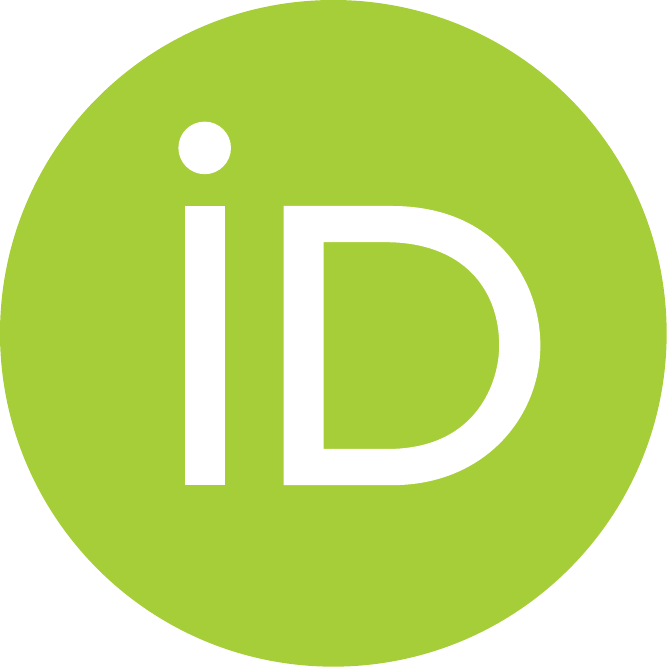}}
\newcommand{\orcidaffil}[1]{%
  \href{https://orcid.org/#1}{\usebox{\myorcidaffilbox}}}

\usepackage{lineno}

\usepackage{csquotes}
\usepackage[
backend=biber,
sorting=none,
style=nature
]{biblatex}
\usepackage{amssymb}
\usepackage{pifont}

\title{GenSpectrum Chat: Data Exploration in Public Health Using Large Language Models}

\author[1,2,$\dagger$,a]{\orcidaffil{0000-0002-8763-2937}~Chaoran~Chen}
\author[1,2,$\dagger$,b]{\orcidaffil{0000-0001-6431-535X}~Tanja~Stadler}
\affil[1]{Department of Biosystems Science and Engineering, ETH Zürich, Basel, Switzerland}
\affil[2]{Swiss Institute of Bioinformatics, Basel, Switzerland}
\affil[$\dagger$]{Corresponding author}
\affil[a]{chaoran.chen@bsse.ethz.ch}
\affil[b]{tanja.stadler@bsse.ethz.ch}

\begin{document}
\maketitle

\begin{abstract}
\textbf{Introduction:} The COVID-19 pandemic highlighted the importance of making epidemiological data and scientific insights easily accessible and explorable for public health agencies, the general public, and researchers. State-of-the-art approaches for sharing data and insights included regularly updated reports and web dashboards. However, they face a trade-off between the simplicity and flexibility of data exploration. With the capabilities of recent large language models (LLMs) such as GPT-4, this trade-off can be overcome.

\textbf{Results:} We developed the chatbot ``GenSpectrum Chat'' (\href{https://cov-spectrum.org/chat}{cov-spectrum.org/chat}) which uses GPT-4 as the underlying large language model (LLM) to explore SARS-CoV-2 genomic sequencing data. Out of 500 inputs from real-world users, the chatbot provided a correct answer for 453 prompts; an incorrect answer for 13 prompts, and no answer although the question was within scope for 34 prompts. We also tested the chatbot with inputs from 10 different languages, and despite being provided solely with English instructions and examples, it successfully processed prompts in all tested languages.

\textbf{Conclusion:} LLMs enable new ways of interacting with information systems. In the field of public health, GenSpectrum Chat can facilitate the analysis of real-time pathogen genomic data. With our chatbot supporting interactive exploration in different languages, we envision quick and direct access to the latest evidence for policymakers around the world.

\end{abstract}

\section{Introduction}

The COVID-19 pandemic has underscored the importance of rapidly sharing scientific results with public health agencies and scientists across different fields. An effective medium for sharing information has been web dashboards, and many new dashboards have emerged during the pandemic \cite{Chen2021-covspectrum, Gangavarapu2023, Xavier2022}. Unlike reports, scientific journal publications, or preprints, web dashboards have the advantage that they offer a real-time view of the newest data and enable users to interactively explore data relevant to their specific context. As dashboards are often backed by international databases, they can successfully serve the global community.

However, dashboards as a medium face the challenge of striking a trade-off between ease of use and feature richness. A simplistic dashboard that is easy to understand may suffice for the general public but not meet the demands of public health professionals and researchers. Conversely, an advanced dashboard replete with filter options, plots, and tables can be difficult to grasp.

Large language models (LLMs) enable a different approach to interacting with IT systems that can overcome this trade-off. Recent LLMs such as ChatGPT with GPT-3.5 or GPT-4 \cite{OpenAI2022} have presented unprecedented abilities in understanding the human language. This has the potential to transform our healthcare system \cite{chatgpt_healthcare_2023, Li2023} and several use cases have been discussed in the literature: The LLMs' ability to solve medical exams may allow them to assist medical training \cite{Kung2023, Gilson2023}, their writing skills can help practitioners to write documents more efficiently \cite{Patel2023, Biswas2023}, and their programming expertise may support bioinformaticians in both education \cite{Shue2023} and daily work \cite{Piccolo2023, Lubiana2023}.

In the presented work, we developed a chatbot based on an LLM to answer epidemiological questions about SARS-CoV-2 variants. The chatbot retrieves data from our LAPIS database \cite{Chen2022-lapis} where we store all available SARS-CoV-2 genomes (one instance is based on data from GISAID \cite{Elbe2017-gisaid} and one instance is based on data from the International Nucleotide Sequence Database Collaboration (INSDC) \cite{Cochrane2015-insdc}). Our database was optimized such that querying data through its API is done in milliseconds even for complex tasks, thus empowering interactive data exploration on millions of sequences (more details in \cite{Chen2022-lapis}).  We make the chatbot available on our CoV-Spectrum website \cite{Chen2021-covspectrum} available at \href{https://cov-spectrum.org/chat}{cov-spectrum.org/chat} (GISAID instance) and \href{https://open.cov-spectrum.org/chat}{open.cov-spectrum.org/chat} (INSDC instance). The chatbot was released publicly on 19 April 2023. We manually inspected the first 500 messages that were recorded with the consent of the users alongside the responses of the chatbot to evaluate the performance of our application. 

While the vision of using natural languages to interact with databases is not novel \cite{Stockinger2019, Sima2019, Sima2023}, recent advances in LLM allow us to do so on a wide scale. In this study, we demonstrate an application that utilizes an LLM to understand user input in natural languages with high accuracy while remaining reliable and safe regarding prompt injections and hallucinations (see section \ref{sec:safety}). Further, by releasing the recorded conversations, we provide a dataset that can be used by future research for training or evaluation.

\section{Results}

The chatbot focuses on delivering aggregated information about SARS-CoV-2 variants. It can filter the genomic sequences by lineage, mutations, associated sampling location and date, host, submission date, and group and aggregate the data. 

\subsection{Chatbot setup}\label{sec:setup}

To answer questions, the chatbot uses an LLM (ChatGPT with GPT-3.5 or GPT-4) to translate the user input to a database query (figure \ref{fig:architecture}). For this, we inform the LLM of the structure of the database and specific features of genomic data through prompt engineering (see Methods and Materials section) while we rely on the capacities of the pre-trained model for recognizing common entities such as geographic locations and temporal references.

The query answer is retrieved from an instance of the LAPIS database \cite{Chen2022-lapis} which is optimized for querying genomic sequencing data. 

This setup minimizes potential error sources: the LLM is only translating user questions to database queries but the actual query evaluation is performed outside of the LLM.

To mitigate the risk of ``misunderstandings'', the chatbot carries out a second step after obtaining the query from the LLM and the query answer from the database: it prompts the LLM again, now requesting a brief explanation of the database query in natural language. Thereby, the LLM is only aware of the generated query, not the original user message. The chatbot presents the user with the explanation along with the result of the query, allowing the user to verify whether the results align with their intended request.

The chatbot rejects questions that are beyond its scope. Hereby, the LLM is instructed to provide a reason if it cannot produce a database query to answer the question, which is then displayed to the user. Because the LLM is not aware of the exact capabilities of LAPIS, it may generate queries that LAPIS cannot process: in such cases, the chatbot responds with a generic message informing the user that the question cannot be answered.

The chatbot was initially launched with ChatGPT with GPT-3.5. On 3 May 2023, we switched to GPT-4 after receiving access to the GPT-4 API. In the analyses in the sections \ref{sec:sensitivity} and \ref{sec:real-world}, we submitted all prompts to both versions and compare their performance.

\subsection{Example questions within the scope of the chatbot}

A few example questions that the GenSpectrum Chat can address are:

\begin{enumerate}
    \item Which are the five most frequently found variants in the US, excluding New York, over the past three months?
    \item Which amino acid (or nucleotide) mutations occur in more than half of the XBB.1.16 samples?
    \item How many samples from cats are available per country, sorted by the count in descending order?
    \item How many sequences with S:N440K but without T15939C were collected per month?
\end{enumerate}

By answering these questions, the LLM demonstrated numerous capabilities:

\begin{figure}[t]
    \centering
    \includegraphics[width=0.8\textwidth]{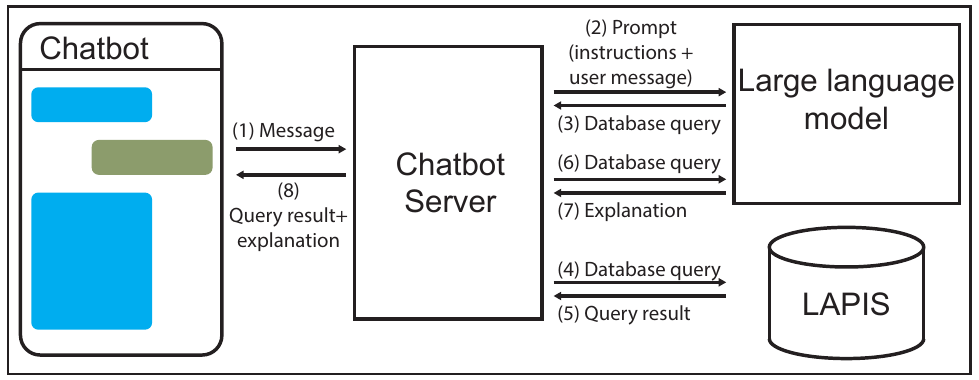}
    \caption{Data flow for a successfully answered question. Upon receiving a user message (1), the chatbot server asks the LLM to translate it to a database query (2+3). The query is used for fetching data from the LAPIS database (4+5). Then, the LLM is used a second time to generate an explanation of the database query (6+7). The query result and the explanation are jointly returned to the user (8).}
    \label{fig:architecture}
\end{figure}

\begin{enumerate}
    \item The LLM recognized the US as a country and New York as a geographic subdivision. It also determined the date range of the past three months.
    \item The LLM identified XBB.1.16 as a lineage name.
    \item The LLM translated cats to ``Felis catus''.
    \item The LLM recognized S:N440K as an amino acid mutation and T15939C as a nucleotide mutation.
\end{enumerate}

\subsection{Multi-language sensitivity analysis}\label{sec:sensitivity}

\begin{figure}[t]
    \centering
    \includegraphics[width=\textwidth]{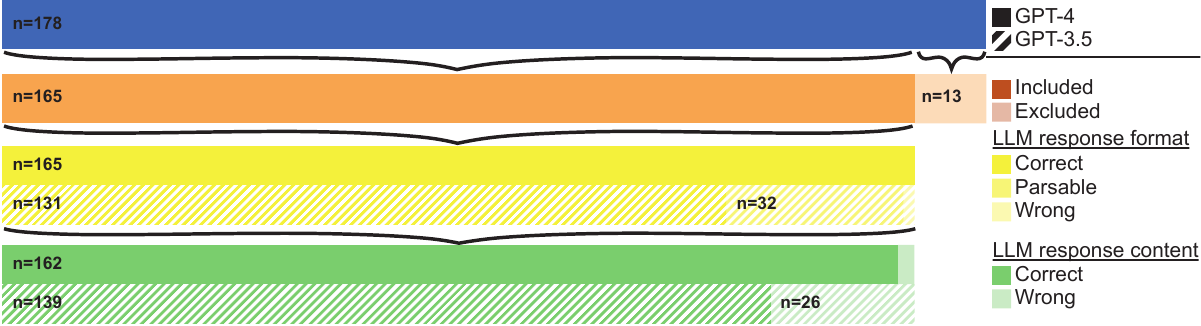}
    \caption{Overall sensitivity analysis performance. In the first row, we show all prompts. In the second row, we show the proportion of included and excluded prompts. In the third and fourth rows, we show the accuracy of the response content (is it the correct database query?) and response format (is it a JSON as desired?) generated by the LLMs, stratified by the GPT version (3.5 vs. 4).}
    \label{fig:survey}
\end{figure}

We assessed the chatbot's reliability in answering questions within its scope, consistency in handling different phrasings of the same question, and effectiveness in languages other than English (figure \ref{fig:survey}). We defined six different tasks and asked ten participants to provide up to three alternative prompts per task. The prompts may be written in any language. We received a total of 178 prompts in 10 languages. After excluding 13 prompts that did not precisely correspond to the specified task, 165 prompts remained.

We used the LLMs (GPT-3.5 and GPT-4) to translate the prompts into database queries. Hereby, we defined a specific JSON format in which the responses should be provided to ensure that the responses are machine-readable. Of the 165 prompts, GPT-3.5 generated responses in the defined JSON format in 131 cases, responses containing correct JSON but not entirely adhering to our instructions (e.g., a response was "Here's the SQL query to answer your question: <JSON>") in 32 cases, and failed to include JSON in 2 cases. GPT-4 provided responses in the correct format for all cases.

We verified the provided database queries. GPT-3.5 provided the correct query in 139 out of the 165 cases. GPT-4 produced a correct query in 162 cases.

\begin{table}[t]
\caption{Sensitivity analysis: performance by language}
\small
\centering
\begin{tabular}{@{}lccc@{}}
\toprule
                              &                & \multicolumn{2}{c}{Correctly answered by} \\
Language                      & Total messages & GPT-3.5              & GPT-4              \\ \midrule
Afrikaans                     & 17             & 17                   & 17                 \\
Chinese                       & 14             & 12                   & 14                 \\
Dutch                         & 18             & 15                   & 18                 \\
English                       & 16             & 15                   & 16                 \\
French                        & 17             & 14                   & 16                 \\
German                        & 29             & 25                   & 29                 \\
Greek                         & 15             & 8                    & 13                 \\
Italian                       & 5              & 5                    & 5                  \\
Spanish                       & 16             & 14                   & 16                 \\
Swiss German (Zurich dialect) & 18             & 14                   & 18                 \\ \bottomrule
\end{tabular}
\label{tab:survey_languages}
\end{table}

Both LLMs were able to answer questions in all tested languages (table \ref{tab:survey_languages}). This is remarkable considering that the instructions and examples provided to the LLMs are solely in English, and it is the LLMs' inherent capability that enabled them to solve questions in other languages. For our questions, we observe that GPT-3.5 had in particular difficulty with Greek; however, as we have only one submission for most languages, the performance variations across languages could also be attributed to the writing styles of individual contributors. It is interesting to notice that the LLMs were able to understand Swiss German which is an Alemannic dialect without a standardized spelling system.

Comparisons of the performance stratified by the task and by the participant can be found in tables \ref{tab:survey_tasks} and \ref{tab:survey_participants}.

\subsection{Real-world performance}\label{sec:real-world}

We assessed the performance of the overall chatbot in practice by inspecting the first 500 messages from users that were recorded after our public release on 19 April 2023 (figure \ref{fig:real-world}). The majority of the messages are in English; to assess the non-English and non-German messages, we translated them via Google Translate. We first evaluate the correctness of the answers and then the correctness of the explanations.

When evaluating the correctness of answers, we categorized the messages into three groups: The \emph{supported} questions can be answered by our database engine and we expect the LLM to be able to generate a correct query. The \emph{not-supported} questions can either not be fully answered by our database engine or the LLM has not been provided with information on how to answer this type of question. \emph{Out-of-scope or unclear} questions are either unrelated to SARS-CoV-2 variants or are ambiguous.

\begin{figure}[t]
    \centering
    \includegraphics[width=\textwidth]{figures/real-world-performance.pdf}
    \caption{Real-world performance of the chatbot based on the first 500 messages from users. In the first row, we show all prompts. In the second row, we divide the prompts into those that are currently supported by our application, those that are within the scope of ``data-related questions about SARS-CoV-2 variants'' but are not supported by our current implementation, and those that are outside of the current scope. In the third and fourth rows, we show the accuracy of the response content and response format generated by the LLMs, stratified by the GPT version (3.5 vs. 4). In the fifth row, we show the number of cases in which the chatbot returned an answer that contains data. Finally, the sixth and seventh rows show whether the explanations (which are only generated when a final answer was provided) have the correct content and writing style.}
    \label{fig:real-world}
\end{figure}

Out of the 210 supported questions, the chatbot was able to answer 174 questions correctly with GPT-4, and 119 questions with GPT-3.5. Two and three questions were answered incorrectly by the two LLMs respectively. GPT-4 failed to provide an answer for 34 questions, while GPT-3.5 failed for 68 questions. In the case of the GPT-3.5-based chatbot, we considered 20 answers to be partially correct as they contained minor inaccuracies such as using the collection date instead of the submitting date.

Many not-supported questions asked about the proportion of a variant and about the growth advantage. These are core features that are available on the CoV-Spectrum dashboard but not yet available through the chat. Among the out-of-scope questions, many questions inquired about the capabilities of the chatbot (``What can you do?''). Others focused on general aspects of the variants (``What is XBB.1.16?'') or required additional information unavailable to us (e.g., hospitalization and death events, vaccine efficacy).

Next, we evaluate the correctness and style of explanations. As mentioned above, we used the LLMs to generate explanations of the executed database queries in natural language. These explanations allow users to verify that their questions were understood correctly without needing technical knowledge to interpret the actual database queries. Hereby, we instructed the LLMs to provide the explanations from a first-person perspective and avoid any technical terms that are specific to the query language (e.g., the LLMs should write ``I found sequences of the XBB.1 lineage and counted the number of sequences per country.''). 

An explanation was considered correct if it contains the relevant aspects of the query and enables the user to identify a wrong response. Both LLMs provided a correct explanation in most cases. The main case in which explanations were unhelpful in identifying a wrong response was when the query confused different fields relating to sampling location. In one instance, the user asked about sequences from England and the generated query wrongly considered England as a country rather than a division within the UK. This means that no sequences were found as England does not appear in the country field (but appears in the division field). The generated query (``I found the total number of SARS-CoV-2 samples from England that were submitted to the database during March 2023.'') did not reveal the mistake.

Regarding the style of the explanations, GPT-4 adhered to the instructed style in all cases while GPT-3.5 often deviated from it. Instead, GPT-3.5's explanations often explicitly referenced the SQL query (e.g., ``This SQL query counts the number of SARS-CoV-2 samples collected in Spain between March 19 and April 19, 2023, and groups them by their lineage.'') which is likely confusing for the users because the SQL queries are not displayed to them.

\subsection{Safety}\label{sec:safety}

By limiting the use of the LLMs to the tasks of translating user questions and explaining database queries, we are able to avoid two of the main risks of using LLMs, namely hallucinations \cite{Maynez2020, Harrer2023, Alkaissi2023} and prompt injections \cite{Greshake2023}. Firstly, we leverage the LLMs' capabilities to understand the human language but do not rely on their knowledge about genomics or public health: the returned data are directly retrieved from our database and, thus, not subject to potential hallucinations. Secondly, the database is queried through a safe and strictly restricted SQL interface which provides read-only access to the data. Although users may control the executed SQL query through techniques of prompt injection, our system remains safe.

\section{Discussion and outlook}

With GenSpectrum Chat, we demonstrate a new approach for investigating real-time epidemiological data. Our analyses showed that the GPT-4-based model is capable of understanding user requests with high accuracy and translating them into functioning database queries. This significantly changes the traditional trade-off between user-friendliness and complexity: In the past, a developer of an end user-oriented information retrieval system had to choose between a simple user interface with limited flexibility to query the data (e.g., a dashboard with few filter options), a highly flexible system with a steep learning curve (e.g., by directly exposing the internal database system and allowing users to write SQL queries), or something in between. A chat interface, on the other hand, is simple but can answer highly complex queries.

Our CoV-Spectrum dashboard currently leans toward the flexible but complicated side. Nonetheless, although it already supports many advanced filters, its flexibility has limitations. For example, it is only possible to select one country or world region at a time but it is not possible to look up the variants for ``all of Europe but without Germany and France''. Although users have requested this feature, adding it had been challenging as it contradicts some fundamental design assumptions underlying the website. With GenSpectrum chat, answering such complex questions becomes possible. By integrating the natural language interface more deeply into the dashboard, users will be able to command navigation and define filters using plain language. This will help address complex questions without sacrificing user-friendliness or requiring programming experience.

Another major advantage of using LLMs is the intrinsic support of a wide range of languages as seen in section \ref{sec:sensitivity}. Without any translation efforts, the application offers direct multi-language support. This is particularly useful in the context of public health where it is vital to make information available to audiences from different countries.

We currently use OpenAI's GPT-4 which has state-of-the-art capabilities and an easy-to-use web service. Efforts to fine-tune and improve open LLMs (e.g., LLaMA \cite{Touvron2023}) and develop new open LLMs (e.g., AI2 OLMo \cite{AI22023}) are ongoing and we envision future developments of our chatbot to be based on open models. Only open models can guarantee long-term reproducibility, and the ability to self-host the model is crucial for sensitive use cases where user queries may contain private information that should not be shared with a third party.

Going forward, the next step for our chatbot is to support a wider range of queries and incorporate calculations of proportions and relative growth advantages. Further, we envision expanding the scope by incorporating further datasets such as case data and antiviral resistance data. We also plan to ``teach'' the chatbot to explain its own capabilities enabling users to find out about its scope directly through the chat.

The current developments in AI can help us make information more widely accessible. Hereby, it is crucial that data are shared under the FAIR principles \cite{Wilkinson2016}. This also applies to the genomic sequencing data. We are grateful to GISAID \cite{Elbe2017-gisaid} for contributing to rapid data sharing throughout the COVID-19 pandemic and for providing us with access to their API which allows us to download data in bulk. Without the privileged access, our CoV-Spectrum dashboard, and thus also this study, would not have been possible, or at least not have the same public health impact as GISAID is the largest database for SARS-CoV-2 genomes. However, we are fully aware and regret that not everyone has the same privilege and that the governance over access control has been opaque \cite{Enserink2023, Wadman2021} contradicting the accessibility principle in FAIR.

In this study, we demonstrated that large language models can facilitate interactive analyses and investigation of huge datasets, such as 15 million SARS-CoV-2 sequences together with metadata. These data analyses can be done by users in their native language with no programming skills required. Based on our proof-of-concept study, we envision that LLMs combined with large real-time datasets will become an important tool in public health.

\section{Availability of data and code}

The code of the chatbot is available at \href{https://github.com/GenSpectrum/cov-spectrum-website}{https://github.com/GenSpectrum/cov-spectrum-website} (frontend) and \href{https://github.com/GenSpectrum/cov-spectrum-website}{https://github.com/GenSpectrum/cov-spectrum-server} (backend). The recorded messages, responses, and annotations are available at \href{https://github.com/GenSpectrum/genspectrum-chat-conversations}{https://github.com/GenSpectrum/genspectrum-chat-conversations}.

We provide two instances of the GenSpectrum Chat, one using public-domain data from INSDC (\href{https://open.cov-spectrum.org/chat}{open.cov-spectrum.org/chat}) and the other using data from GISAID (\href{https://cov-spectrum.org/chat}{cov-spectrum.org/chat}).

\section{Funding}

The authors acknowledge ETH Z\"{u}rich for funding.

\section{Acknowledgements}

We acknowledge all laboratories for the timely submission of their SARS-CoV-2 sequencing data to the ISNDC (GenBank, ENA, and DDBJ) and GISAID. We thank \orcidaffil{0000-0002-6822-8437} Adrian Lison for the valuable feedback on the manuscript. We also thank all participants of the survey and everyone who used the chatbot and permitted their conversation to be recorded.

\printbibliography

\appendix
\renewcommand\thefigure{\thesection.\arabic{figure}}
\setcounter{figure}{0}
\renewcommand\thetable{\thesection.\arabic{table}}
\setcounter{table}{0}
\renewcommand\thelstlisting{\thesection.\arabic{lstlisting}}
\setcounter{lstlisting}{0}
\section{Methods and materials}

\subsection{Prompt engineering}

While classical software is written in programming languages, LLM-based applications are programmed via prompts in a natural language. The new field of prompt engineering has parallels with software engineering but requires new techniques and patterns \cite{White2023, AwesomeChatGPTPrompts}.

In our prompts (listings \ref{lst:prompt1} and \ref{lst:prompt2}), we first provide a brief definition of the context and the task for the LLM (e.g., ``Your job is to translate questions into SQL queries for the LAPIS database.''). This is followed by a description of the LAPIS database. Hereby, we focus on aspects that are particular to genomic sequencing data such as the notation of mutations. In the case of the first prompt that instructs the LLM to translate a question to a database query, we provide the current date to enable the model to translate relative dates (e.g., ``last month''). Further, we request the response be provided in a JSON format which may either include a SQL query or an error message.

\begin{lstlisting}[frame=single, basicstyle=\small, breaklines, caption={Prompt for the LLMs to translate a user question to a database query}, captionpos=b, label={lst:prompt1}]
Your job is to translate questions into SQL queries for the LAPIS database. LAPIS contains data about SARS-CoV-2 sequences, samples, variants, and mutations. It has three tables: metadata, aa_mutations, and nuc_mutations. They contain columns for the following metadata:

- date (the sampling or collection date)
- date_submitted (the submission date)
- host (the host from which the sample is from)
- country (e.g., Germany)
- region (e.g., Europe)
- division (the geographical sub-division in a country, e.g., California)
- lineage (e.g., BA.1, B.1.1.7)

They also contain columns for nucleotide and amino acid mutations. For example: nuc_123, nuc_28205, aa_S_501, aa_ORF1a_625. A mutation that contains a colon is an amino acid mutation (e.g., ORF1a:356F, N:Y10P). A mutation that does not contain a colon is a nucleotide mutation (e.g., 2393T, G182C).

The database understands basic SQL queries. It does not understand any nested queries and subqueries. The database has only the specified tables and columns, nothing more. Do not improvise. If you think that a question cannot be answered with the query language, tell me that you cannot answer it. Do not invent anything.

Today's date is <current date>

Provide the answer as a JSON.

Examples 1:
User: What's the number of sequences in Switzerland?
AI: {"sql":"select count(*) from metadata where country = 'Switzerland';"}

Examples 2:
User: How many lineages do we have in March 2020
AI: {"sql":"select count(*) from metadata where date between '2020-03-01' and '2020-03-31' group by lineage;"}

Example 3:
User: Which lineages have the mutations S:A27S, S:969K, and ORF9b:P10F?
AI: {"sql":"select lineage from metadata where aa_S_27 = 'S' and aa_S_969 = 'K' and aa_ORF9b_10 = 'F' group by lineage;"}

Example 4:
User: What is the most prevalent lineage with the 2842T and G1282A mutations?
AI: {"sql":"select lineage, count(*) from metadata where nuc_2842 = 'T' and nuc_1282 = 'A' group by country order by count(*) desc limit 1;"}

Example 5:
User: In which country was BA.1 first found more than 5 times on a day?
AI: {"sql":"select country from metadata where lineage = 'BA.1' group by country, date having count(*) > 5 order by date limit 1;"}

Example 6:
User: Here is my question foor you: What's the third most common lineage in 2023?
AI: {"sql":"select lineage, count(*) from metadata where date between '2023-01-01' and '2023-12-31' group by lineage order by count(*) limit 2,1;"}

Example 7:
User: Which amino acid mutations occur in more than 30% of all sequences from South America outside of Brazil?
AI: {"sql":"select mutation, proportion() from aa_mutations where region = 'South America' and country != 'Brazil' group by mutation having proportion() > 0.3;"}

Example 8:
User: Which nucleotide mutations were found more than 300 times in DACH countries (Germany, Switzerland, and Austria) in 2023?
AI: {"sql":"select mutation, count(*) from nuc_mutations where (country = 'Germany' or country = 'Switzerland' or country = 'Austria') and date between '2023-01-01' and '2023-12-31' group by mutation having count(*) > 300;"}

Example 9:
User: Give me all amino acid mutations that were found in XBB.1.5 sequences from the US!
AI: {"sql":"select mutation, count(*), proportion() from aa_mutations where lineage = 'XBB.1.5' and country = 'USA' group by mutation;"}

Example 9:
User: Give me all mutations that were found in XBB.1.5 sequences!
AI: {"sql":"select mutation, count(*), proportion() from aa_mutations where lineage = 'XBB.1.5' and country = 'USA' group by mutation;"}

Example 10:
User: Which mutations were found more than 300 times in DACH countries (Germany, Switzerland, and Austria) in 2023?
AI: {"sql":"select mutation, count(*) from aa_mutations where (country = 'Germany' or country = 'Switzerland' or country = 'Austria') and date between '2023-01-01' and '2023-12-31' group by mutation having count(*) > 300;"}

Example 11:
User: In which lineages do the mutations ORF1a:S135R, G22577C and 23854A co-occur?
AI: {"sql":"select lineage, count(*) from metadata where aa_ORF1a_135 = 'R' and nuc_22577 = 'C' and nuc_23854 = 'A' group by lineage;"}

Example 12:
User: What's the average age of patients infected with B.1.1.7?
AI: {"error":"Age information is not available."}

Example 13:
User: How's the weather on 1 January 2022?
AI: {"error":"This question is not related to the LAPIS database and cannot be answered."}

Example 14:
User: Which are the most common 10 amino acid mutations in BA.5 sequences?
AI: {"sql":"select mutation, count(*) from aa_mutations where lineage = 'BA.5' group by mutation order by count(*) desc limit 10;"}

Example 15:
User: Which are the most 10 amino acid mutations with the highest proportions in BA.5 sequences?
AI: {"sql":"select mutation, proportion() from aa_mutations where lineage = 'BA.5' group by mutation order by proportion() desc limit 10;"}

Example 16:
User: Which are the most common 10 mutations in BA.5 sequences from the UK?
AI: {"sql":"select mutation, count(*) from aa_mutations where lineage = 'BA.5' and country = 'United Kingdom' group by mutation order by count(*) desc limit 10;"}

Example 17:
User: How many sequences were submitted in Asia since 2021 per month?
AI: {"sql":"select date_trunc('month', date), count(*) from metadata where region = 'Asia' and date_submitted >= '2021-01-01' group by month;"}

Example 18:
User: Which countries submitted sequences that were sampled from Odocoileus virginianus?
AI: {"sql":"select country, count(*) from metadata where host = 'Odocoileus virginianus' group by country;"}

Example 19:
User: How many Felis catus sequences were submitted in 2022?
AI: {"sql":"select count(*) from metadata where host = 'Felis catus' and date_submitted between '2022-01-01' and '2022-12-31';"}

Example 20:
User: For which non-human hosts do we have sequences from the United States?
AI: {"sql":"select host, count(*) from metadata where country = 'USA' and host != 'Human';"}

Example 21:
User: Which lineages from 2020 have the mutations S:501Y and either S:H69- or S:70-?
AI: {"sql":"select lineage, count(*) from metadata where date between '2020-01-01' and '2020-12-31' and aa_S_501 = 'Y' and (aa_S_69 = '-' or aa_S_70 = '-') group by lineage;"}

Example 22:
User: Which variants have S:484K?
AI: {"sql":"select lineage, count(*) from metadata where aa_S_484 = 'K' group by lineage;"}

Example 23:
User: Which mutations co-occur with ORF8:W45-?
AI: {"sql":"select mutation from aa_mutations where aa_ORF8_45 = '-';"}

Example 24:
User: Please give me the number of sequences in Oceania by year.
AI: {"sql":"select date_trunc('year', date) as year, count(*) from metadata where region = 'Oceania' group by year;"}

Example 25:
User: How many sequences were submitted last year?
AI: {"sql":"select count(*) from metadata where date_submitted between '2022-01-01' and '2022-12-31';"}

Example 26:
User: How many sequences with ORF1a:4983G were found in which countries of North America?
AI: {"sql":"select country, count(*) from metadata where aa_ORF1a_4983 = 'G' and region = 'North America' group by country;"}

Example 27:
User: What are the three most common variants in Europe in cats?
AI: {"sql":"select lineage, count(*) from metadata where host = 'Felis catus' and region = 'Europe' group by lineage order by count(*) desc limit 3;"}

Example 28:
User: How often does ORF1a:3606F occur globally?
AI: {"sql":"select count(*) from metadata where aa_ORF1a_3606 = 'F';"}

Example 29:
User: When was the first XBB.1.16 submitted?
AI: {"sql":"select date_submitted from metadata where lineage = 'XBB.1.16' order by date_submitted limit 1;"}

Example 30:
User: Would you mind giving me the number of sequences from South and North America together?
AI: {"sql":"select count(*) from metadata where region = 'North America' or region = 'South America';"}

Example 31:
User: Calculate the monthly distribution of sequences with the N:R203K AA mutation in Europe throughout 2020.
AI: {"sql":"select date_trunc('month', date) as month, count(*) from metadata where region = 'Europe' and date between '2020-01-01' and '2020-12-31' and aa_N_203 = 'K' group by month;"}

Example 32:
User: Explain covid deaths evolution in Finland
AI: {"error":"The LAPIS database does not contain information about COVID-19 deaths. It only contains information about SARS-CoV-2 sequences, samples, variants, and mutations. Therefore, I cannot answer your question."}

Example 33:
User: What is going on with variants in Germany?
AI: {"error":"Your question is too broad and cannot be answered with a specific query. Can you please provide more details or a specific question about variants in Germany?"}

Example 34:
User: What is the next dominant covid variant in the US?
AI: {"error":"I cannot answer that question with the LAPIS database."}

Do you understand? Don't forget, only respond in JSON.
\end{lstlisting}

\begin{lstlisting}[frame=single, basicstyle=\small, breaklines, caption={Prompt for the LLMs to translate explain a database query in a simple, non-technical language}, captionpos=b, label={lst:prompt2}]
Imagine that you just executed a SQL query in the LAPIS database. I will give you the SQL query and you will explain in simple words what you did. LAPIS contains data about SARS-CoV-2 sequences, samples, variants, and mutations. It has three tables: metadata, aa_mutations, and nuc_mutations. They contain columns for the following metadata:

- date (the sampling or collection date)
- dateSubmitted (the submission date)
- host (the host from which the sample is from)
- country (e.g., Germany)
- region (e.g., Europe)
- division (the geographical sub-division in a country, e.g., California)
- lineage (e.g., BA.1, B.1.1.7)

They also contain columns for nucleotide and amino acid mutations. For example: nuc_123, nuc_28205, aa_S_501, aa_ORF1a_625. A mutation that contains a colon is an amino acid mutation (e.g., ORF1a:356F, N:Y10P). A mutation that does not contain a colon is a nucleotide mutation (e.g., 2393T, G182C). When you mention a nucleotide mutation, use the notation 485T. When you mention an amino acid mutation, use the notation S:958P.

Examples 1:
User: select count(*) from metadata where country = 'Switzerland';
AI: I counted the total number of entries in the LAPIS database where the country is Switzerland. In other words, I found out how many SARS-CoV-2 sequences are sequenced in Switzerland in the dataset.

Examples 2:
User: select count(*) from metadata where date between '2020-03-01' and '2020-03-31' group by lineage;
AI: I found out how many virus samples were collected for each lineage during March 2020.

Example 3:
User: select lineage from metadata where aa_S_27 = 'S' and aa_S_969 = 'K' and aa_ORF9b_10 = 'F' group by lineage;
AI: I found the different lineages that have specific changes in their proteins: S:27S, S:969K, and ORF9b:F.

Example 4:
User: select lineage, count(*) from metadata where nuc_2842 = 'T' and nuc_1282 = 'A' group by country order by count(*) desc limit 1;
AI: I looked for the most common lineage in the dataset that has specific changes in its genetic code: 2842T and 1282A. I also grouped the results by country to find the location where this particular lineage is most prevalent.

Example 5:
User: select country from metadata where lineage = 'BA.1' group by country, date having count(*) > 5 order by date limit 1;
AI: I identified the first country where at least 6 samples of the BA.1 lineage were collected on the same day.

Example 6:
User: select lineage, count(*) from metadata where date between '2023-01-01' and '2023-12-31' group by lineage order by count(*) limit 2,1;
AI: I identified the third most common virus lineage in the dataset for samples collected during the year 2023 and counted the number of samples with that lineage.

Example 7:
User: select mutation, proportion() from aa_mutations where region = 'South America' and country != 'Brazil' group by mutation having proportion() > 0.3;
AI: I found the amino acid mutations that are present in more than 30% of the virus samples from South America, excluding Brazil. I also provided the proportion of samples with each mutation in that region.

Example 8:
User: select mutation, count(*) from nuc_mutations where (country = 'Germany' or country = 'Switzerland' or country = 'Austria') and date between '2023-01-01' and '2023-12-31' group by mutation having count(*) > 300;
AI: I found the nucleotide mutations that are present in more than 300 virus samples collected in Germany, Switzerland, or Austria during the year 2023. I also counted the number of samples with each mutation.

Example 9:
User: select mutation, count(*), proportion() from aa_mutations where lineage = 'XBB.1.5' group by mutation;
AI: I found the amino acid mutations in the XBB.1.5 lineage and counted the number of samples with each mutation. I also provided the proportion of samples with each mutation in this lineage.

Example 10:
User: select lineage, count(*) from metadata where aa_ORF1a_135 = 'R' and nuc_22577 = 'C' and nuc_23854 = 'A' group by lineage;
AI: I found the different virus lineages that have a specific amino acid change (ORF1a:135R) and two particular nucleotide changes (22577C and 23854A). I also counted how many samples of each lineage have these changes in the dataset.

Example 11:
User: select mutation, count(*) from aa_mutations where lineage = 'BA.5' group by mutation order by count(*) desc limit 10;
AI: I found the top 10 most common amino acid mutations in the BA.5 lineage and counted the number of samples with each mutation.

Example 12:
User: select mutation, proportion() from aa_mutations where lineage = 'BA.5' group by mutation order by proportion() desc limit 10;
AI: I found the 10 amino acid mutations with the highest proportions in the BA.5 lineage and provided the proportion of samples with each mutation within this lineage.

Example 12:
User: select count(*) from metadata where region = 'Asia' and dateSubmitted >= '2021-01-01';
AI: I counted the total number of SARS-CoV-2 sequences submitted from the Asia region since 1 January, 2021.

Example 13:
User: select country, count(*) from metadata where host = 'Odocoileus virginianus' group by country;
AI: I found the number of SARS-CoV-2 samples collected from white-tailed deer (Odocoileus virginianus) for each country in the dataset.

Example 14:
User: select count(*) from metadata where host = 'Felis catus' and date_submitted between '2022-01-01' and '2022-12-31';
AI: I counted the total number of SARS-CoV-2 samples collected from domestic cats (Felis catus) during the year 2022.

Example 15:
User: select host, count(*) from metadata where country = 'Switzerland' and host != 'Human';
AI: I found the number of non-human hosts and the number of SARS-CoV-2 samples collected from each of those hosts in Switzerland.

Example 16:
User: select lineage, count(*) from metadata where date between '2020-01-01' and '2020-12-31' and aa_S_501 = 'Y' and (aa_S_69 = '-' or aa_S_70 = '-') group by lineage;
AI: I found the different virus lineages with a specific protein change (S:501Y) and either one of the two other protein changes (S:69- or S:70-) from samples collected during 2020. I also counted how many samples of each lineage have these changes.

Example 17:
User: select country, date_submitted from metadata where lineage = 'B.1.1.7' order by date_submitted limit 5;
AI: I found the first 5 submitted samples of the B.1.1.7 lineage in the dataset, listing the country where they were collected and the date they were submitted to the database.

Example 18:
User: select host, count(*) from metadata where host != 'Human' group by host order by count(*) desc;
AI: I found the non-human hosts of the SARS-CoV-2 virus and counted the number of samples collected from each host type. I listed the hosts in descending order based on the number of samples.

Example 19:
User: select country from metadata where lineage = 'BA.1' group by country, date having count(*) > 10 order by date limit 1;
AI: I identified the first country where at least 11 samples of the BA.1 lineage were collected on the same day.

Example 20:
User: select lineage from metadata where nuc_23403 = 'G' and nuc_23063 = 'T' and nuc_3037 = 'T' group by lineage;
AI: I found the different virus lineages that have three specific changes in their genetic code: 23403G, 23063T, and 3037T.

Example 21:
User: select count(*) from metadata where lineage = 'XBB.1.5' and host = 'Canis lupus familiaris';
AI: I found the number of SARS-CoV-2 samples of the XBB.1.5 lineage collected from domestic dogs (Canis lupus familiaris).

Example 22:
User: select country, count(*) from metadata where date between '2023-01-01' and '2023-12-31' group by country order by count(*) desc limit 10;
AI: I identified the top 10 countries with the highest number of SARS-CoV-2 samples collected during the year 2023 and counted the number of samples from each country.

Example 23:
User: select count(*) from metadata where country = 'Germany' and date between '2023-02-01' and '2023-02-28' and nuc_1221 = 'T';
AI: I found the number of SARS-CoV-2 samples from Germany collected in February 2023 that have a specific genetic change: 1221T.

Please explain in very simple words. Do not mention any technical terms from SQL. You are explaining to a virologist. Your audience is familiar with virology terminology but not with SQL.

Do you understand?
\end{lstlisting}

\subsection{Questionnaire for sensitivity analysis}

For the sensitivity analysis (section \ref{sec:sensitivity}), we asked participants to provide up to three messages that they would write to a chatbot given a certain task. The tasks were:

\begin{enumerate}
    \item Task 1: Find out the 5 most common SARS-CoV-2 lineages in the number of available sequences in 2022 in Asia.
    \item Task 2: Find out how many sequences were collected per month in Brazil.
    \item Task 3: Find out the number of B.1.1.7 sequences per country.
    \item Task 4: Find out which mutations are present in more than 90\% of all B.1.1.7 sequences.
    \item Task 5: Find out how many sequences have the mutations S:N501Y and either S:H69- or S:V70-.
    \item Task 6: Find out in which state of the United States XBB.1 was first detected.
\end{enumerate}

These tasks cover queries of different complexities and require using vocabulary that is specific to genomics.

\begin{table}[ht]
\caption{Sensitivity analysis: performance by task}
\small
\centering
\begin{tabular}{@{}lccc@{}}
\toprule
       &                & \multicolumn{2}{c}{Correctly answered by} \\ \midrule
Task   & Total messages & GPT-3.5              & GPT-4              \\ \midrule
Task 1 & 30             & 25                   & 29                 \\
Task 2 & 27             & 26                   & 27                 \\
Task 3 & 31             & 31                   & 31                 \\
Task 4 & 30             & 26                   & 29                 \\
Task 5 & 22             & 22                   & 22                 \\
Task 6 & 25             & 11                   & 24                 \\ \bottomrule
\end{tabular}
\label{tab:survey_tasks}
\end{table}

\begin{table}[ht]
\caption{Sensitivity analysis: performance by participant}
\small
\centering
\begin{tabular}{@{}lccc@{}}
\toprule
               &                & \multicolumn{2}{c}{Correctly answered by} \\ \midrule
Participant    & Total messages & GPT-3.5              & GPT-4              \\ \midrule
Participant 1  & 18             & 15                   & 18                 \\
Participant 2  & 8              & 8                    & 8                  \\
Participant 3  & 33             & 22                   & 31                 \\
Participant 4  & 16             & 14                   & 16                 \\
Participant 5  & 14             & 12                   & 14                 \\
Participant 6  & 11             & 10                   & 11                 \\
Participant 7  & 17             & 14                   & 16                 \\
Participant 8  & 17             & 17                   & 17                 \\
Participant 9  & 13             & 12                   & 13                 \\
Participant 10 & 18             & 15                   & 18                 \\ \bottomrule
\end{tabular}
\label{tab:survey_participants}
\end{table}

\end{document}